\documentclass[seceq]{ptptex}
\usepackage{amssymb,amsmath,epsfig}

\title{Interpretation of the Five Dimensional
Quantum Propagation of a Spinless Massless Particle }

\author{Romulus \textsc{Breban}\footnote{Present address: 
Department of Biomathematics, University of California, Los Angeles, 90024.}}

\inst{ Institute for Research in Electronics and Applied Physics and Department 
of Physics, University of Maryland, College Park, Maryland 20740.}

\abst{We consider a five dimensional (5D) space-time with a space-like fifth
dimension. We implement a quantum formalism by path integrals, and postulate
that all the physical information on a 5D massless particle propagation is 
provided by the statistics over null paths in this 5D space-time. If the 5D 
metric is independent of the fifth coordinate, then the propagation problem 
can be reduced to four dimensions by foliation along the fifth coordinate, and
we obtain a formulation of 4D Quantum Mechanics. If the 5D metric is 
independent of time, we foliate along the time coordinate, and obtain a 
formulation of 4D Statistical Mechanics. If the 5D metric is independent of 
both time and the fifth coordinate, then Quantum and Statistical Mechanics are
pictures of the same 5D reality. We also discuss the foliation of a 
proper space dimension, the Klein-Gordon equation, and a 5D Special Relativity,
completing our interpretation of the 5D geometry.}

\begin{document}
\maketitle

PACS: 03.65.-w, 05.30.-d, 04.50.+h, 31.15.Kb

Keywords: quantum mechanics, quantum statistical mechanics, Kaluza-Klein 
gravity, path integrals

\newpage

\section{Introduction}
\label{sec:intro}

In 1921, Kaluza \cite{Kaluza21_} proposed a five dimensional (5D) 
geometrical framework for the unification of the gravitational and the 
electromagnetic interactions. Following the model of the 4D General 
Relativity, Kaluza discussed both the field equations and the particle 
propagation along 5D time-like geodesics, trying to establish the role of the
fifth dimension for the traditional 4D physics. Kaluza \cite{Kaluza21_}
proposed that the fifth dimension be proportional to electric charge.
Several years later, Klein 
studied quantum aspects of the 5D propagation \cite{Klein26,Klein27}, and
noticed that the assumption of a compact and planckian fifth dimension leads
to the quantization of electric charge, while addressing the question of why 
the fifth dimension is not experimentally observable \cite{Klein27}. The
Kaluza-Klein approach to the field theory has later been devoted an extensive
amount of literature \cite{Wesson,many}; not the same has happened
to the theory of the 5D particle propagation, in spite of its fundamental 
importance. In fact, Kaluza \cite{Kaluza21_}
immediately recognized that his 5D particle propagation theory cannot apply
to the electron and other elementary particles \cite{Einstein}. In 1984,
Gegenberg and Kunstatter \cite{Gegenberg} were led to the conclusion that, 
indeed, it is seemingly impossible to naturally describe the propagation of
charged light particles (i.e., with rest mass less than the Planck mass)
within a classical or quantum 5D Kaluza-Klein framework. In String Theory
this problem seems solved by giving up the concept of point particle. Here
we ask ourselves whether this problem is deeply rooted in the Kaluza-Klein
interpretation of the 5D geometry, and search for alternatives.

In the attempt to construct a {\it scale-invariant gravity}, Wesson 
\cite{Wesson,Wess83} proposed an interpretation of a 5D geometry where the
fifth dimension, not necessarily compact, is proportional to 
mass (i.e., $x^5=Gm/c^2$, where $G$ is the Newton constant, and $c$ is the
velocity of light). This idea is known in the literature as 
{\it Kaluza-Klein gravity}. Investigating 5D time-like geodesics in this new
framework, Wesson was led to predict the existence of an exotic fifth force,
yet to be discovered experimentally.
Recently, Seahra and Wesson \cite{Wess01} approached the problem of the 
anomalous fifth force for null geodesics. Their study is limited to 5D
manifolds with metrics conformally conjugated to that of the Kaluza-Klein 
theory in the absence of electromagnetic fields. [However, they do not 
require the fifth dimension be compact.] Furthermore, the metric
depends on the fifth coordinate {\it exclusively} through the conformal factor
\cite{kkmetric} which is required to depend {\it only} on the fifth 
coordinate. Then, writing the 4D {\it on-shell} constraint 
$p^\mu p_\mu=-m^2c^2$ [$\mu=0,1,2,3$, and the 4D metric is
$diag(-1,1,1,1)$] as $p^Ap_A=0$ ($A=0,1,2,3,5$), where $p^5=p_5=mc$, inspired
Seahra and Wesson to pursue and demonstrate the idea that a 5D massless 
particle propagating along a 5D null geodesic can be seen as a 4D massive 
particle propagating along a 4D time-like geodesic. Then, they 
show that for the 5D null geodesic propagation, the anomalous fifth force can
be removed by a reparametrization, which in turn introduces ambiguities
in defining the 4D proper time of the corresponding 4D time-like geodesic.
Thus, it seems that Seahra and Wesson have found an appropriate, anomaly-free
4D interpretation of the 5D null geodesics in the absence of electromagnetic
fields (and imposing other constraints mentioned above). However, a close look
to their theory, reveals several consistency problems. First, the fact that 
the fifth dimension is proportional to mass is inconsistent with the 5D
interpretation of the on-shell constraint. Consider the simple case of a 5D 
massless particle following a null geodesic in a flat, Minkovski-like 5D 
metric. Assume that the particle has constant 4D mass. By the interpretation 
of the fifth dimension, the particle is moving in a hyperplane of constant 
$x^5$. By the interpretation of the 4D on-shell constraint, the particle has 
constant momentum along the fifth dimension, and thus, unless $p^5=0$ (i.e., 
$m=0$, or the particle is a 4D photon), the fifth coordinate changes with time.
Second, choosing a particular proper time for the 5D null geodesics as seen in
four dimensions breaks 
5D covariance, and this is inconsistent with the claim of a 5D gravity. While 
it is remarkable that 5D null geodesics can be regarded as 4D time-like 
geodesics, it is not clear {\it why} an observer would use such an 
interpretation.\\

In the attempt to find the proper relation between the 4D and the 5D 
particle propagation, this paper discusses a new interpretation of the 5D 
space-time geometry. We give a new 4D interpretation to 5D null propagation, 
and we apply it to the case of weak fields.
To address the propagation problem in the presence of 
electromagnetic fields, in both classical and
quantum regimes, we equip the 5D geometry with a quantum principle for 5D 
spinless and massless particle propagation. The geodesic propagation is then 
obtained by taking the classical limit.

An important idea for our work is the distinction between {\it active},
{\it pasive} and {\it inertial} mass \cite{Jamm61}. In short, active mass is
the source of gravitational field, and passive mass is the object the
gravitational field acts upon \cite{Wessmass}. We adopt the Equivalence 
Principle of General Relativity which identifies the inertial and the passive 
mass. The equivalence
between the passive and the active mass is required by the Action-Reaction
Principle of Newtonian Mechanics. In General Relativity, we may consider
the active and the passive mass as distinct concepts, whose comparison is not
necessarily meaningful \cite{FN8}.  We also distinguish
active and passive electric charge.

Another important concept for our work is that of path integral introduced
to physics by Feynman \cite{Feyn48,Feyn65,Feyn72} in 1948. 
Here we consider path integrals to be coordinate-free quantities. Given
two points 1 and 2 in a space-time manifold, a sum over all paths from 1 to 2
depends on the choice of the points 1 and 2, but it is independent of 
coordinates. However, in the practice of calculating path integrals, global
coordinates as a consistent labeling of all points in the space-time manifold
are extremely useful. In 4D quantum physics, path integrals are now very 
powerful tools. Feynman \cite{Feyn48,Feyn65,Feyn72} rewrote the quantum 
principles of 4D Quantum and Statistical Mechanics in
the same language of path integrals. In this paper, we implement a 5D path 
integral quantum principle to describe the massless and spinless 5D 
propagation. We then show that 4D Quantum and Statistical Mechanics can emerge
as different particular interpretations of the 5D geometry endowed with this
Feynman quantum principle. Thus, in five dimensions, we obtain not only the 
unification of electromagnetic and gravitational interactions, but 
also a unification of 4D Quantum and 4D Statistical Mechanics. 
Perhaps, this is not surprising given that 4D mechanics and 4D statistics have
been developing now for many decades side by side, borrowing each other ideas 
and formalism both in the quantum mechanical and in the quantum field 
theoretical frameworks, and being just a Wick rotation away.\\

We start with a 5D space-time having a space-like fifth dimension.
We do not request the fifth dimension be compact, and, in 5D context,
we do not restrict the transformations of coordinates to be cylindrical
[i.e., $y^\mu=y^\mu(x^\nu)$, $\mu,\nu,... =0,1,2,3$, and $y^5=y^5(x^5)$]. 
Thus, in principle,
the fifth dimension is observable both in the field equations and particle
propagation. It is the purpose of this paper to discuss how the fifth
dimension is revealed to an electrically uncharged 4D observer 
that perceives geometrically only the first four dimensions, but not the
fifth. In four dimensions, the fifth dimension is not manifested as a 
geometrical entity, but rather through its consequences \cite{pigeon}. 

Consider the situation where a 5D spinless and massless quantum particle, 
called a {\it 5D photon}, propagates in a 5D curved space-time. We  
formulate the setup of the propagation problem introducing a path
integral quantum principle.
{\it A zero rest mass particle is created at a point 1 of a 5D space-time with
fixed metric, and then annihilated at a point 2 in the future cone of 1. 
We assume that the existence of this particle does not alter the space-time
geometry. Maximal information on the particle's quantum propagation is 
obtained from the number of all null paths between points 1 and 2.}
We call this setup {\it 5D Quantum Optics}. 
 We discuss breaking of 5D covariance when a 4D observer measures a 5D photon
 and assigns it a 4D physical picture for 5D metrics independent of some 
coordinates. 
In the case of translational symmetry along the fifth
coordinate, our procedure has formal similarities to the traditional Kaluza-Klein
dimensional reduction.  In contrast, the spontaneous compactification
of String Theory is a proper physical phenomenon; a macroscopic 
observer detects all 11 (or 10) dimensions of the space-time up to quantum effects due
to the small diameters of the compact dimensions.

The outline of the paper is as follows. We proceed with presenting the general
 case of the 5D Quantum Optics (Sec.~\ref{sec:QOptics}). In Section 
\ref{sec:QM}, we introduce the particular 
situation where the 5D metric is independent of the fifth space-like 
coordinate. We first shortly discuss the field equations. Then, in 
Sec.~\ref{sec:MCanQM}, we show that the 5D null path integral is equivalent to
a 4D path integral over time-like paths with specified length. Thus, we 
construct a microcanonical ensemble for our 4D Quantum Mechanics from the 
microcanonical ensemble of the 5D Quantum Optics. In Sec.~\ref{sec:CanQM} 
we introduce the corresponding 4D canonical ensemble \cite{Brody}. 
Then, we make the connection between the nonrelativistic limit of our 
formulation of Quantum Mechanics and Feynman's. We conclude that our 
Quantum Mechanics is anomaly-free and equivalent to the traditional Feyman 
formulation. In Sec.~\ref{sec:SM}, we investigate another 
particular case of the 5D Quantum Optics which we interpret as a formulation 
of Statistical Mechanics. The state of thermal equilibrium implies that 
macroscopic observables do not change in time. Presumably, this can be 
achieved for a 5D metric which is time-independent. In this case, we 
reformulate the 5D Quantum Optics as a Statistical Mechanics for a single 
spinless particle which belongs to an ensemble of non-interacting particles. 
It seems unavoidable that our path integral formulation of Statistical 
Mechanics needs the Ergodic Principle in order to establish relations with 
experiment.  Also, our Statistical Mechanics describes only ensembles of 
discernable particles, unless correction factors are introduced `by hand'. If 
the 5D metric is independent of both time and the fifth coordinate, Quantum 
and Statistical Mechanics make different pictures of the same 5D reality.
In Sec.~\ref{sec:KG}, we present how the Klein-Gordon equation
appears from the 5D formalism. Section \ref{sec:5DST} discusses the main
aspects of the 5D Special Relativity, and makes further connections with
traditional 4D physics. In Sec.~\ref{sec:x3} we discuss the case where
the 5D metric is independent on a proper space dimension (i.e., $x^3$) and 
then conclude our work.

\section{5D Quantum Optics}
\label{sec:QOptics}

We consider a 5D space-time with metric $h_{AB}$, ($A,B,... =0,1,2,3,5$)
having the signature $diag(-1,1,1,1,1)$. All transformations of
coordinates are allowed, like in the case of 4D gravity. [We do not restrict
to considering only cylindrical transformations of coordinates -- i.e.,
$y^\mu=y^\mu(x^\nu)$, $\mu,\nu,... =0,1,2,3$, and $y^5=y^5(x^5)$.] 

Consider a 5D massless particle propagation between two measurement events 1 
and 2 (with 2 in the future cone of 1) in the 5D space-time. We implement a 
quantum formalism by path integrals and postulate that all physical 
information about the particle propagation can be obtained from the number of 
all null paths between 1 and 2
\begin{eqnarray}
{\cal R}(2,1)=\sum_{
 \genfrac{}{}{0pt}{3}{\mbox{\scriptsize\mbox{all null paths}}}
 {\mbox{\scriptsize\mbox{between 1 and 2}}}
}1\equiv \int_{\begin{array}{c}
\genfrac{}{}{0pt}{3} {\mbox{\scriptsize\mbox{all null paths}}}
 {\mbox{\scriptsize\mbox{between 1 and 2}}}
\end{array}}[d^5x].\nonumber
\end{eqnarray}
${\cal R}(2,1)$ assumes equal probability of realization for every 5D
null path between 1 and 2, and will be treated like a microcanonical sum 
\cite{FN}. We notice that ${\cal R}(2,1)$ is positively defined and 
conformally invariant. It also satisfies the following selfconsistency 
relation which results from its geometrical meaning \cite{jacobian}
\begin{eqnarray}
{\cal R}(2,1)=\int d^5x^{(3)}\sqrt{|h|}\; {\cal R}(2,3){\cal R}(3,1).
\nonumber
\end{eqnarray}
By $\int d^5x^{(3)}\sqrt{|h|}$, we denote the volume integration over all
points 3 in the future cone of 1 and in the past cone of 2.
The points 1 and 2 would be special {\it observable} events of creation
 and annihilation of a
particle, but in this theory we do not request them to be geometrical
elements in space-time for the simplicity of the 5D geometry.
If the points 1 and 2 have no geometrical meaning, ${\cal R}(2,1)$
is not invariant to an arbitrary change of coordinates.

\section{Quantum Mechanics}
\label{sec:QM}
We assume that $h_{AB}$ does not depend on $x^5$ and
foliate the 5D space-time along the fifth coordinate. That is, we see the
static 5D space-time as a 4D space-time {\it evolving} in the fifth
coordinate \cite{Misn73}
\begin{eqnarray}
\label{eq:foliation}
ds_5^2=h^{AB}dx_A dx_B=g^{\mu \nu}\left(dx_\mu
+N_\mu dx_5 \right) \left(dx_\nu+N_\nu dx_5 \right)+N^2dx_5^2.\nonumber
\end{eqnarray}
$N$ is the lapse, and $N_\mu$ is the shift of the 4D foliation. 
In matrix form, we have
\begin{eqnarray}
h^{AB}=\left(\begin{array}{cc} g^{\mu \nu} & N^\mu \\
                                   N^\nu  &  N^\rho N_\rho +N^2
\end{array} \right) \Leftrightarrow
h_{AB}=\left(\begin{array}{cc}
g_{\mu \nu}+\frac{N_\mu N_\nu}{N^2} & -\frac{N_\mu}{N^2} \\
              -\frac{N_\nu}{N^2}  & \frac{1}{N^2}
\end{array}\right).\nonumber
\end{eqnarray}
With the notation
\begin{eqnarray}
\label{eq:nt}
 N_\mu&=&-\frac{q}{c^2}A_\mu, \nonumber\\
\frac{1}{N}&=&\Phi, \nonumber
\end{eqnarray}
 where $q$ is a parameter to be discussed in Sec.~\ref{sec:CanQM},
we get the so-called Kaluza-Klein parameterization of the 5D metric
\begin{eqnarray}
\label{eq:fol1}
h_{AB}=\left(\begin{array}{cc}
g_{\mu \nu}+\frac{q^2}{c^4}\Phi^2 A_\mu A_\nu & \frac{q}{c^2}\Phi^2 A_\mu \\
              \frac{q}{c^2}\Phi^2 A_\nu   & \Phi^2
\end{array}\right).
\end{eqnarray}\\

 We now proceed to discuss the field equations. The case where the 5D 
space-time manifold is not Ricci flat introduces a 5D energy-momentum
tensor which has to be motivated from the experimental point of view.
In this paper we restrict to 5D Ricci flat space-time
manifolds; i.e., where $R_{AB}^5=0$. 
The field equations for the Kaluza-Klein parametrization of the
$x^5$-independent metric [given by Eq.~\eqref{eq:fol1}], and some of their 
exact solutions can be found in Ref.~\cite{Wesson}.
From our subsequent results, the {\it physical} 4D metric providing
the trajectories of particles is not $g_{\mu\nu}$, but
$\tilde{g}_{\mu\nu}=g_{\mu\nu}/\Phi^2$. Anticipating this, we are
interested to see how the field equations look under the
aforementioned 4D conformal transformation. Straightforward calculations
yield
\begin{eqnarray}
\label{eq:fieldeq2}
\tilde{R}^4_{\mu \nu}-\frac{1}{2} \tilde{g}_{\mu
\nu}\tilde{R}^4&=&\frac{q^2}{2c^4}\tilde{T}^{EM}_{\mu \nu}+
\tilde{T}_{\mu\nu}^\Phi,\nonumber\\
\tilde{\bigtriangledown}^\mu \tilde{F}_{\mu \nu}&=&
-3\frac{\tilde{\bigtriangledown}^\mu \Phi}
{\Phi}\tilde F_{\mu \nu},\\
\frac{1}{\Phi}\tilde{\square} \Phi
&=&\frac{q^2}{4c^4}\tilde{F}_{\mu \nu}\tilde{F}^{\mu \nu}
+\frac{1}{6}\tilde{T}_{\mu\nu}^\Phi\tilde{g}^{\mu\nu},\nonumber
\end{eqnarray}
where $\tilde{R}^4_{\mu \nu}$ and $\tilde{\bigtriangledown}_\mu$ are, 
respectively, the Ricci tensor and the covariant derivative of the metric 
$\tilde{g}_{\mu\nu}$, 
$\tilde F_{\mu\nu}=\partial_\mu A_\nu-\partial_\nu A_\mu$, and
\begin{eqnarray}
\tilde{\bigtriangledown}^\mu\equiv\tilde{g}^{\mu\nu}
\tilde{\bigtriangledown}_\nu,\nonumber\quad
\tilde{\square}\equiv\tilde{g}^{\mu\nu} \tilde{\bigtriangledown}_\mu
\tilde{\bigtriangledown}_\nu,\nonumber\quad
\tilde{F}^{\mu \nu}\equiv\tilde{F}_{\alpha \beta}\;
\tilde{g}^{\mu\alpha}\tilde{g}^{\nu\beta},\nonumber
\end{eqnarray}
\begin{eqnarray}
\tilde T^{EM}_{\alpha \beta}=\tilde g_{\beta\mu}
\tilde F_{\alpha\lambda}\tilde F^{\lambda\mu}+
\frac{1}{4}\tilde g_{\alpha\beta}\tilde F_{\mu\lambda}\tilde 
F^{\mu\lambda},\nonumber
\end{eqnarray}
\begin{eqnarray}
\tilde{T}_{\mu\nu}^\Phi\equiv\frac{1}{\Phi}
\left[\tilde{\bigtriangledown}_\mu \tilde{\bigtriangledown}_\nu \Phi
-\frac{2}{\Phi}(\tilde{\bigtriangledown}_\mu\Phi)
(\tilde{\bigtriangledown}_\nu\Phi)\right]
-\frac{1}{\Phi}\tilde{g}_{\mu\nu}\tilde{g}^{\alpha\beta}
\left[\tilde{\bigtriangledown}_\alpha \tilde{\bigtriangledown}_\beta \Phi
-\frac{2}{\Phi}(\tilde{\bigtriangledown}_\alpha\Phi)
(\tilde{\bigtriangledown}_\beta\Phi)\right].
\nonumber
\end{eqnarray}
Equations \eqref{eq:fieldeq2} represent the Einstein and the Maxwell equations
 in an interpretation of {\it induced} matter \cite{Kaluza21}. The empty 5D 
space-time
seems to contain active matter when it is given 4D physical meaning. $A_\mu$,
which so far was just an abstract field proportional to the shift of the 4D
foliation, becomes now the electromagnetic field {\it on} a 4D manifold with 
the metric $\tilde g_{\mu\nu}$.  The terms in the RHS of the Einstein equations
are interpreted as induced sources of gravitational field, and the 
RHS of the Maxwell equations is interpreted as induced 4-current density of 
{\it active} electrical charge. Under the 4D conformal transformation, the 
sources of gravitational field completely separate into electromagnetic 
sources and $\Phi$-sources; this is not the case for the untransformed field 
equations (see \cite{Wesson}). The term $\tilde{T}_{\mu\nu}^\Phi$ 
depends exclusively on $\Phi$ and $g_{\mu\nu}$, it is independent of $A_\mu$, 
and characterizes the distribution of induced active matter in the 4D foliated
 manifold. 
 
 In traditional 5D Kaluza-Klein theory, the
parameter $q$ would be chosen $q=4c^2\sqrt{\pi G}$ such that the above Einstein 
equations are similar to their 4D formulation. Here instead, we will use the 
equations of particle propagation (see Sec.~\ref{sec:CanQM}) to interpret the
parameter $q$ as {\it passive} specific electrical charge. 

Equations \eqref{eq:fieldeq2} and, consequently, their 4D interpretation are not
invariant to a general 5D transformation of coordinates. For them to hold, we
restrict to cylindrical transformations (i.e., $y^\mu=y^\mu(x^\nu)$, and 
$y^5=y^5(x^5)$ \cite{cyl}); noncylindrical ones possibly {\it mix} the 
gravitational and the electromagnetic fields. However, consider the situation 
of a 4D observer in a frame where Eqs.~\eqref{eq:fieldeq2} represent an 
adequate description of reality. It is not obvious how the 4D observer would 
perform a certain noncylindrical transformation of coordinates since this 
transformation essentially belongs to a 5D geometry. For the 4D observer, 
noncylindrical transformations are not just plain 4D geometrical 
transformations like those of 
the General Relativity with the metric $\tilde g_{\mu\nu}$.\\ 

We now give 4D physical interpretation to the sum over 5D null paths 
introduced in Sec.~\ref{sec:QOptics}. This interpretation  also breaks 5D 
covariance, and will not hold if all possible transformations of coordinates 
are applied;  we again restrict to cylindrical transformations. Consider
the 5D null path element 
\begin{eqnarray}
\label{eq:5Dnulltau}
ds_5^2=h_{AB}dx^Adx^B
=g_{\mu \nu} dx^\mu dx^\nu+\Phi^2 \left(dx^5+\frac{q}{c^2}A_\mu
dx^\mu \right)^2=0, 
\end{eqnarray}
and solve for $dx^5$ 
\begin{eqnarray}
\label{eq:nul5D}
dx^5= \pm \frac{1}{\Phi}\sqrt{-g_{\mu \nu}dx^\mu dx^\nu}-
\frac{q}{c^2}A_\rho dx^\rho
\equiv \pm \sqrt{-\tilde{g}_{\mu \nu}dx^\mu dx^\nu}-
\frac{q}{c^2}A_\rho dx^\rho.
\end{eqnarray}
 We note that if $dx^5=0$ and $q=0$ ($q$ has meaning of specific charge; see 
Sec.~\ref{sec:CanQM}), Eq.~\eqref{eq:nul5D} yields 4D null paths describing 
4D photons.

If $\tilde g_{\mu\nu}$ and $A_\mu$ are independent of $x^5$, integrating 
\eqref{eq:nul5D} between two events 1 and 2 of the 5D space-time yields
\begin{eqnarray}
\label{eq:q}
\int_1^2 dx^5=\Delta x^5=\int_1^2 \left( \pm
\sqrt{-\tilde{g}_{\mu \nu}dx^\mu dx^\nu}-\frac{q}{c^2}A_\rho dx^\rho \right)
\equiv D^{21}_{\pm}.
\end{eqnarray}
Notice that 5D covariance is lost in Eq.~\eqref{eq:q}. The RHS of \eqref{eq:q} 
belongs exclusively to a 4D space-time manifold for which we define the 
infinitesimal time-like distance as
\begin{eqnarray}
ds_{4\;\pm}=\pm\sqrt{-\tilde{g}_{\mu \nu}dx^\mu dx^\nu}-\frac{q}{c^2}
A_\rho dx^\rho.\nonumber
\end{eqnarray}
The LHS of \eqref{eq:q} is a constant that depends only on the initial
and final points of the particle propagation. Therefore, counting null paths
between events 1 and 2 in the 5D manifold is equivalent in this case with
counting 4D time-like paths of length $D^{21}_\pm=\Delta x^5$ between the 4D 
projections $1^4$ and $2^4$ of the 5D events 1 and 2, respectively.

\subsection{Microcanonical Ensemble of Quantum Mechanics}
\label{sec:MCanQM}

\begin{figure}\begin{center}
\mbox{\epsfig{file=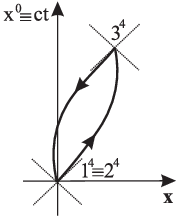,width=60mm}}
\caption{Closed loop in 4D space-time: {\it Quantum loop}. The dotted lines
represent the local causal cones of $1^4$ and $3^4$.}
\label{fig:fig2}
\end{center}\end{figure}

The 5D null path integral ${\cal R}(2,1)$ is equivalent to the following 4D
path integral over time-like paths with length $D=\Delta x^5$
\begin{eqnarray}
R_{\pm}(D,2^4,1^4)=\sum_{
\genfrac{}{}{0pt}{3}{\mbox{\scriptsize\mbox{all time-like paths of}}}
 {\mbox{\scriptsize\mbox{length {\em D} between $1^4$ and $2^4$}}}
}1\equiv \int_{D_{\pm}^{21}=D}[d^4x]. \nonumber
\end{eqnarray}
$R_{\pm}(D,2^4,1^4)$ is positively defined and satisfies a
selfconsistency relation resulting from its geometrical meaning
\begin{eqnarray}
\label{eq:cr}
R_{\pm}(D,2^4,1^4)=\int d^4x^{(3)}\sqrt{|\tilde{g}|}
\int_{-\infty}^{\infty} dD' \;R_{\pm}(D',2^4,3^4)R_{\pm}(D-D',3^4,1^4).
\end{eqnarray}
Event $2^4$ must be in the future cone of $1^4$. By $3^4$ we denoted any
intermediary 4D event which is in the future cone of $1^4$ and in the past 
cone of $2^4$. By the notation $\int d^4x^{(3)}$ we understand integrating
over the set of all points $3^4$ satisfying the aforementioned condition.
Consistency with Eq.~\eqref{eq:q} demands that the distance along a path 
between $1^4$ and $2^4$ in a future cone is minus the distance along the same 
path between $2^4$ and $1^4$ in a past cone
\begin{eqnarray}
\label{eq:qdist}
D_{\pm}^{21}=-D_{\pm}^{12}.
\end{eqnarray}
We call a 4D loop made from one path in future cone and another in past cone
a {\it quantum loop}; see Fig.~\ref{fig:fig2}. The microcanonical sum over 
quantum loops, $R_{\pm}(D,1^4,1^4)$, is invariant to gauge transformations of 
$A_\mu$ (see Sec.~\ref{sec:CanQM}), and plays a special role in this formalism.

\subsection{Canonical Ensemble of Quantum Mechanics}
\label{sec:CanQM}
Not only that path integrals are hard to calculate, but constrained path 
integrals must be even harder. We eliminate the constraint in 
$R_{\pm}(D,2^4,1^4)$ by a Fourier transform with respect to $D$
\begin{eqnarray}
\label{eq:can}
K_{\pm}(\lambda^{-1},2^4,1^4)=\int^{\infty}_{-\infty}dD \;
e^{iD\lambda^{-1}}R_{\pm}(D,2^4,1^4)
=\int [d^4x]\; e^{iD_{\pm}^{21}\lambda^{-1}}.\nonumber
\end{eqnarray}
This Fourier transform is justified by the translation symmetry of the 5D 
space-time manifold along the fifth coordinate. Writing $D_{\pm}^{21}$ 
explicitly [see Eq.~\eqref{eq:q}] yields
\begin{eqnarray}
\label{eq:can+}
K_{\pm}(\lambda^{-1},2^4,1^4)=\int [d^4x]
\exp \left[ {i\lambda^{-1}\int_1^2 \left( \pm
\sqrt{-\tilde{g}_{\mu \nu}dx^\mu dx^\nu}-\frac{q}{c^2}A_\rho dx^\rho \right)}
 \right].
\end{eqnarray}
$K_{\pm}$ provides a complete quantum description of a 5D particle 
propagation, as
seen in four dimensions. Performing a Fourier transformation in both hands of 
\eqref{eq:cr} yields a selfconsistency relation for $K_{\pm}$
\begin{eqnarray}
\label{eq:cr2}
K_{\pm}(\lambda^{-1},2^4,1^4)=\int d^4x^{(3)} \sqrt{|\tilde{g}|} \; K_{\pm}
(\lambda^{-1},2^4,3^4) \;K_{\pm}(\lambda^{-1},3^4,1^4).\nonumber
\end{eqnarray}\\

We now proceed to compare the nonrelativistic limit of this path integral 
formulation of Quantum Mechanics to that proposed by Feynman 
\cite{Feyn48,Feyn65}. For doing so, we assume that our 5D manifold has the 
topology of the flat space-time, and that $x^A$ are pseudocartesian 
coordinates (i.e., all coordinates 
$x^A$ take values on the whole real axis). Also, we consider that the metric
$\tilde{g}_{\mu \nu}$ describes a weak gravitational field ($i,j=1,2,3$)
\begin{eqnarray}
\label{eq:nnm}
\tilde{g}_{\mu \nu}= \left(\begin{array}{cc}
-1-\frac{2V}{mc^2} & 0 \\ 0  & \delta_{ij}\end{array}\right),\nonumber
\end{eqnarray}
where $|2V/(mc^2)|\ll 1$, $V$ corresponding to the Newtonian gravitational
potential ($V\propto m$). We expand the nonrelativistic limit of
$D_-\lambda^{-1}$ in $2V/(mc^2)$, and keep only the first order
\begin{eqnarray}
\label{eq:lD}
D_-^{21}\lambda^{-1}&\approx&-\lambda^{-1}c(t_2-t_1)
\nonumber\\&+&\lambda^{-1}c \int^2_1 dt \left[
\frac{1}{2c^2}\left(\frac{d\overrightarrow{x}}{dt}\right)^2
-\frac{q}{c^3}\overrightarrow{A}\frac{d\overrightarrow{x}}{dt}
-\frac{q}{c^2}A_0-\frac{V}{mc^2}\right],
\end{eqnarray}
where $c\,t\equiv x^0$. If we choose $\lambda$ to be the Compton wavelength
of a particle with (passive) mass $m$ (i.e., $\lambda^{-1}=mc/\hbar$), we 
obtain that the RHS of \eqref{eq:lD} is
$[-mc^2(t_2-t_1)+S(m,mq)]/\hbar$, where by 
$S(m,mq)$ we denote the traditional nonrelativistic mechanical action of a 
particle with mass $m$ and electrical charge $mq$. This identification gives 
physical meaning to our abstract formalism. Thus, $\lambda ^{-1}$ is 
essentially the passive rest mass of a 4D particle \cite{m=0}, and $q$ is the 
particle's passive specific electrical charge. $D_-$ (i.e., $\Delta x^5$) is 
essentially the particle's mechanical action, and conjugated by Fourier 
transform to $\lambda ^{-1}$. Similar calculations yield 
$D_+\lambda^{-1}\approx [mc^2(t_2-t_1)-S(m,-mq)]/\hbar$. Thus, if we say that 
$D_-\lambda^{-1}$ describes
a particle propagating forward in time, then $D_+\lambda^{-1}$ describes an
antiparticle propagating backwards in time. Note that $D_+\lambda^{-1}$ can 
also be written as $D_+\lambda^{-1}\approx[-(-m)c^2(t_2-t_1)+S(-m,mq)]/\hbar$,
and interpreted as describing a particle with mass $-m$ and charge $mq$ 
propagating forward in time.

We now recognize that the nonrelativistic limit of $K_{\pm}$ is proportional 
to a Feynman path integral. We can show, following the same steps as Feynman 
\cite{Feyn48,Feyn65}, that the nonrelativistic limit of $K_{\pm}$ is the 
propagator of a Schr\"odinger equation
\begin{eqnarray}
\pm\frac{\hbar}{i}\frac{\partial K_\pm}{\partial t}=\frac{1}{2m}
\left[\frac{\hbar}{i}\nabla-\frac{mq}{c}\overrightarrow{A}\right]
^2K_\pm\mp(mq)A_0K_\pm+ VK_\pm+ mc^2K_\pm.\nonumber
\end{eqnarray}
We emphasize that this Schr\"odinger equation results from the metric 
$\tilde{g}_{\mu\nu}$, and not from the original 4D metric $g_{\mu\nu}$. The 
 fields observable by the propagation of a 4D quantum particle are only
$\tilde{g}_{\mu\nu}$ and $A_\mu$. $\Phi$ is a scalar field unobservable in 
Quantum and Classical Mechanics.\\

We now resume the analysis of the relativistic theory. Consider a gauge 
transformation of $A_\rho$ which must not change the observables of the 
canonical ensemble
\begin{eqnarray}
A_\rho=A'_\rho+\partial_\rho A.\nonumber
\end{eqnarray}
From \eqref{eq:can+}, this gauge transformation implies a local $U(1)$ 
transformation of $K_{\pm}$
\begin{eqnarray}
K_{\pm}(\lambda^{-1},2^4,1^4)=\exp\left\{- i\lambda^{-1}
\frac{q}{c^2}[A(2^4)-A(1^4)]\right\}
\nonumber\\ \times
\int [d^4x]\; \exp \left[ {i\lambda^{-1}\int_1^2 \left( \pm
\sqrt{-\tilde{g}_{\mu \nu}dx^\mu dx^\nu}-\frac{q}{c^2}A'_\rho dx^\rho
\right)} \right].\nonumber
\end{eqnarray}
Equivalently, we can write a 5D noncylindrical coordinate transformation such 
that the transformed 5D metric contains the gauge transformed electromagnetic 
field $A'_\rho$ [see Eq.~\eqref{eq:5Dnulltau}]
\begin{eqnarray}
\label{eq:noncyl}
y^\mu&=&x^\mu, \\
y^5&=&x^5-\frac{q}{c^2}A(x^\mu)+(const.)\,.\nonumber
\end{eqnarray}
As a result of this noncylindrical transformation, the new mechanical action 
${D'}_{\pm}^{21}=\Delta y^5$ relates to the old mechanical action 
$D_{\pm}^{21}=\Delta x^5$ as
\begin{eqnarray}
D_{\pm}^{21}={D'}_{\pm}^{21}-\frac{q}{c^2}[A(2^4)-A(1^4)].\nonumber
\end{eqnarray}
Thus, the actions of the gauge transformation of $A_\rho$, of the local $U(1)$ 
transformation, and of the noncylindrical change of coordinates 
\eqref{eq:noncyl} on $K_\pm$ are equivalent.

In general, $D_{\pm}^{21}$ is not gauge invariant, but can be made gauge 
invariant
if the 5D events 1 and 2 have the same 4D projection (i.e., if $1^4= 2^4$). 
In particular, $D_{\pm}^{21}$ is gauge invariant if its integration path is a 
quantum loop. Thus, the canonical sum over quantum loops is gauge invariant,
and can be written as
\cite{fn}
\begin{eqnarray}
K_{\pm}(\lambda^{-1},1^4,1^4)=\int d^4x^{(3)} \sqrt{|\tilde{g}|} \;
K_{\pm}(\lambda^{-1},3^4,1^4) \; K_{\pm}(\lambda^{-1},1^4,3^4),\nonumber
\end{eqnarray}
or using \eqref{eq:qdist},
\begin{eqnarray}
\label{eq:core}
K_{\pm}(\lambda^{-1},1^4,1^4)=\int d^4x^{(3)}
\sqrt{|\tilde{g}|} \; K_{\pm}(\lambda^{-1},3^4,1^4) \;
K^*_{\pm}(\lambda^{-1},3^4,1^4),
\end{eqnarray}
where * is symbol for complex conjugation. The quantum kernel (or the 
propagator) $K_{\pm}(\lambda^{-1},1^4,1^4)$ from one point of the 4D 
space-time to itself is normalizable to 1. Equation \eqref{eq:core} represents
a formula for the quantum probability, and stands at the very core of Quantum 
Mechanics. It can be further generalized for wavefunctions. The generalization
 is rather straightforward, and we will not present it here. See Feynman 
\cite{Feyn48,Feyn65} for a discussion on how to construct mathematical 
formulae holding for wavefunctions when formulae for
the quantum kernel are provided.\\

We now shortly discuss the classical limit of the path integral formulation in
 the canonical ensemble. The most substantial contribution to the canonical 
sum over paths is given by the paths with stationary length. Such paths are 
the trajectories of Classical Mechanics satisfying Hamilton's Principle
\begin{eqnarray}
\label{eq:st}
\delta D_{\pm}=0.\nonumber
\end{eqnarray}
We consider $ D_{\pm}$ to be the action of a massive particle propagating in a
4D manifold with the metric $\tilde g_{\mu\nu}$. A suitable choice of 
parametrization for the geodesics as observed in four dimensions is
\begin{eqnarray}
d\tau^2=-\tilde{g}_{\mu\nu}dx^\mu dx^\nu.\nonumber
\end{eqnarray}
With this parametrization, the Euler-Lagrange equations turn out to be
\begin{eqnarray}
\label{eq:EL}
\dot{x}^\mu\tilde{\bigtriangledown}_\mu \dot{x}^\tau\pm
\frac{q}{c^2}\tilde{g}^{\tau\sigma}F_{\sigma\rho}\dot{x}^\rho=0,~\mbox{where }
\dot{x}^\mu=\frac{dx^\mu}{d\tau},
\end{eqnarray}
which are the expected classical equations of motion of a massive particle 
with specific charge $q$, in gravitational and electromagnetic fields. If we 
rewrite \eqref{eq:5Dnulltau} as
\begin{eqnarray}
\label{eq:other5D}
\left(\dot{x}^5+\frac{q}{c^2}A_\mu\dot{x}^\mu \right)^2=1,
\end{eqnarray}
then \eqref{eq:EL} and \eqref{eq:other5D} represent the 5D null geodesic
equations of the metric $h_{AB}$ written in a parametrization that breaks 5D
covariance. A 5D massless particle propagation corresponds to a 4D massive 
particle propagation.

The 5D manifold has a Killing field $\xi^A=(0,0,0,0,1)$ due to the invariance
 of the metric $h_{AB}/\Phi^2$ to translations along $x^5$. Thus, there is a
conserved quantity along a 5D null geodesic
\begin{eqnarray}
\bar m\equiv\frac{h_{AB}}{\Phi^2}\,\xi^A\frac{dx^B}{d\sigma}=
\frac{dx_5}{d\sigma},\nonumber
\end{eqnarray}
with $\sigma$ being the affine parameter of the 5D null geodesic of the metric
$h_{AB}/\Phi^2$. $\bar m$ can be thought as the fifth component of the 5D 
photon's 5-momentum. We associate the existence of $\bar m$ to the fact that 
the passive mass is a constant of motion in 4D Classical Mechanics.

\section{Statistical Mechanics}
\label{sec:SM}

We now assume that the metric $h_{AB}$ is independent of $x^0$. Following the 
same steps of the foliation along $x^5$, we parameterize the 5D metric similar
 to the Kaluza-Klein form
\begin{eqnarray}
\label{eq:fol2}
h_{AB}=\left(\begin{array}{cc}
                      -\phi^2 & -\frac{Q}{c^2}\phi^2 a_\nu \\
-\frac{Q}{c^2}\phi^2 a_\mu   & G_{\mu \nu}-\frac{Q^2}{c^4}\phi^2 a_\mu a_\nu
\end{array}\right),
\end{eqnarray}
where now $\mu,\nu,...=1,2,3,5.$ We can get the field equations for the case 
of a 5D Ricci flat manifold in this parameterization of the 5D metric by the 
formal substitution $A_\mu\rightarrow a_\mu$, 
$\tilde G_{\mu\nu}\rightarrow \tilde g_{\mu\nu}$ and 
$\Phi^2\rightarrow -\phi^2$ in Eqs.~\eqref{eq:fieldeq2}. We note 
that $a_\mu$ is similar to the electromagnetic field $A_\mu$.\\

We look for an interpretation of the sum over null paths. 
Solving the null path element in $dx^0$ and then integrating yields
\begin{eqnarray}
\label{eq:sd}
\Delta x^0\equiv\int_1^2 dx^0=
\int_1^2 \,\left(\pm \frac{1}{\phi}\sqrt{G_{\mu \nu}dx^\mu dx^\nu}-
\frac{Q}{c^2}a_\rho dx^\rho\right)\equiv d^{21}_{\pm}.
\end{eqnarray}
$\Delta x^0$ plays role of action in Statistical Mechanics. Therefore, in this 
context, $x^0$ cannot be interpreted as coordinate time. We think of it as 
{\it physical time}, intrinsically related to the propagation phenomenon as 
mechanical action relates to the particle propagation in Quantum Mechanics.
An essential difference from the case of Quantum Mechanics is that every 
observable phenomenon takes a positive amount of time; i.e., $\Delta x^0> 0$. 
Mathematically, we
request that $\Delta x^0$, as a functional of paths, is bounded from below 
\cite{fnmin}. This implies that we have only one choice of distance on 
the foliated 4D manifold [c.f., Eq.~\eqref{eq:sd}]
\begin{eqnarray}
dS_{4\;+}=\frac{1}{|\phi|}
\sqrt{G_{\mu \nu}dx^\mu dx^\nu}-\frac{Q}{c^2}a_\rho dx^\rho=
\sqrt{\tilde{G}_{\mu \nu}dx^\mu dx^\nu}-
\frac{Q}{c^2}a_\rho dx^\rho.\nonumber
\end{eqnarray}
The observable 4D metric in Statistical Mechanics is not $G_{\mu\nu}$, but
$\tilde{G}_{\mu \nu}\equiv G_{\mu\nu}/\phi^2$. $\phi$ is unobservable in 
Statistical Mechanics, as $\Phi$ is unobservable in Quantum Mechanics. Denote 
the projections of the 5D events 1 and 2 to the 4D Rienmannian manifold by 
$1^4$ and $2^4$, respectively. Similar to the case of Quantum Mechanics, the 
${\cal R}(2,1)$ sum over 5D null paths is equivalent to a sum over paths from 
$1^4$ to $2^4$ in a 4D manifold, having a certain length $d_+^{21}=d$. For the
 Statistical Mechanics interpretation to hold, we restrict to cylindrical 
transformations of coordinates determined by the foliation along $x^0$ 
[i.e., $y^\mu=y^\mu(x^\nu)$, and $y^0=y^0(x^0)$]. We formally keep the range 
of values for the lengths of paths, 
$d$, to be the whole real axis, instead we consider that all functions of $d$ 
vanish under the integral for values of the argument lower than a certain 
threshold $d_{\rm min}$.

\subsection{Microcanonical Ensemble of Statistical Mechanics}

${\cal R}(2,1)$ is equivalent to a path integral in a 4D Rienmannian manifold
with the infinitesimal distance $dS_{4+}$
\begin{eqnarray}
\rho_{+}(d,2^4,1^4)=\sum_{
\genfrac{}{}{0pt}{3}{\mbox{\scriptsize\mbox{all paths of length d}}}
 {\mbox{\scriptsize\mbox{between $1^4$ and $2^4$}}}
}1\equiv \int_{d_+=d}[d^4x].\nonumber
\end{eqnarray}
$\rho_{+}(d,2^4,1^4)$ is positively defined and must satisfy the following
selfconsistency relation deriving from its geometrical meaning
\begin{eqnarray}
\label{eq:crs}
\rho_{+}(d,2^4,1^4)=\int d^4x^{(3)}\sqrt{|\tilde{G}|}
\int_{-\infty}^{\infty} dd' \;  \rho_{+}(d',2^4,3^4)\rho_{+}(d-d',3^4,1^4).
\end{eqnarray}
Event $3^4$ is not constrained by causal structure in this case.
We define the thermodynamic entropy as
\begin{eqnarray}
S_{+}(G_{\mu\nu}, d)=k_B\ln\int d^4x\sqrt{|\tilde G|} \; \rho_{+}(d,1^4,1^4),\nonumber
\end{eqnarray}
where $k_B$ is the Boltzmann constant. $S_{+}(G_{\mu\nu},d)$ is 4D covariant, 
and invariant to gauge transformations of $a_\mu$.

\subsection{Canonical Ensemble of Statistical Mechanics}
\label{sec:CanSM}

We introduce the canonical ensemble quite similarly to the case of Quantum 
Mechanics. However, since $d_+$ is bounded from below, a Laplace transform 
will be more appropriate than a Fourier transform 
\begin{gather}
\label{eq:scan}
k_{+}(\Lambda^{-1},2^4,1^4)=\int^{\infty}_{-\infty}dd \;
e^{-d\Lambda^{-1}}\rho_{+}(d,2^4,1^4)\nonumber\\
=\int [d^4x]\exp \left[ {-\Lambda^{-1}\int_1^2 \left(
\sqrt{\tilde{G}_{\mu \nu}dx^\mu dx^\nu}- \frac{Q}{c^2}a_\rho dx^\rho
\right)} \right].\nonumber
\end{gather}
We expect that $\Lambda^{-1}$ denotes a physical concept characterizing 
thermal equilibrium. We get a selfconsistency relation for $k_{+}$ by applying
 a Laplace transform on both sides of \eqref{eq:crs}
\begin{eqnarray}
\label{eq:smcank}
k_{+}(\lambda^{-1},2^4,1^4)=\int d^4x^{(3)} \sqrt{|\tilde{G}|} \;
k_{+}(\lambda^{-1},2^4,3^4) \; k_{+}(\lambda^{-1},3^4,1^4).\nonumber
\end{eqnarray}

Similar to the case of Quantum Mechanics, we find a noncylindrical 
transformation of coordinates which is equivalent to a gauge transformation of
 $a_\rho$. If the points 1 and 2 have the same 4D projection, then
$k_{+}(\Lambda^{-1},1^4,2^4=1^4)$ (i.e., the path integral over 4D loops) is 
invariant to gauge transformations of $a_\rho$. We call a 4D loop a {\it 
statistical loop}. By definition, a statistical loop indicates {\it 
Equilibrium} Statistical Mechanics. In 5D, the particle is measured at 
different moments of time 1 and 2 to be at the same 4D position. The 
definition of the Massieu function for the canonical ensemble is
\begin{eqnarray}
\Psi_{+}(G_{\mu\nu},\Lambda^{-1})=k_B\ln \int d^4x\sqrt{|\tilde G|} \;
k_{+}(\Lambda^{-1},1^4,1^4).\nonumber
\end{eqnarray}
$\Psi_{+}(G_{\mu\nu},\Lambda^{-1})$ is 4D covariant, and invariant to gauge
transformations of $a_\mu$.\\

We now proceed to make the connection with Feynman's formulation of 
Statistical 
Mechanics. We first shortly review Feynman's version \cite{Feyn65,Feyn72}.
Consider a time independent problem of Nonrelativistic Quantum Mechanics for a 
single spinless particle. If $E_a$ and $\Phi_a$ are the eigenvalues and the 
corresponding eigenvectors of the Hamiltonian operator, then the quantum 
kernel between two points $1^4$ and $2^4$ (with $2^4$ in the future cone of 
$1^4$) can be written as
\begin{eqnarray}
K(\lambda^{-1};
\overrightarrow{x}_2,t_2;\overrightarrow{x}_1,t_1)=
\sum_a \Phi_a^*(\overrightarrow{x}_2)\Phi_a(\overrightarrow{x}_1)e^{i
E_a(t_2-t_1)/\hbar}.\nonumber
\end{eqnarray}
On the other hand, from the traditional formalism of Statistical Mechanics,
the canonical density of states for one particle is
\begin{eqnarray}
\label{eq:Toll}
\hat{k}(\beta;\overrightarrow{x})=\sum_a \Phi_a^*(\overrightarrow{x})
\Phi_a(\overrightarrow{x})e^{-\beta E_a},
\end{eqnarray}
where $\beta=1/(k_BT)$ with $T$ being Kelvin temperature. The striking 
similarity of the last two formulae suggested Feynman that they may come from 
similar formalisms. He wrote $\hat{k}$ as
\begin{eqnarray}
\hat k(\overrightarrow{x}_2,u_2;\overrightarrow{x}_1,u_1)=\sum_a
\Phi_a^*(\overrightarrow{x_2})
\Phi_a(\overrightarrow{x_1})e^{-E_a(u_2-u_1)/\hbar},\nonumber
\end{eqnarray}
where $u_2-u_1\equiv\beta\hbar$; $u$ is a new coordinate with dimension of 
time and physical meaning of temperature to minus one. It was then easy to guess
a path integral formulation for Statistical Mechanics \cite{FN2}
\begin{eqnarray}
\hat k(\overrightarrow{x}_2,\beta\hbar;\overrightarrow{x}_1,0)=
\int [d^3x]\exp \left\{ -\frac{1}{\hbar}\int_0^{\beta\hbar}du
\left[\frac{m}{2}\left|\frac{d\overrightarrow{x}}{du} \right|^2+
v(\overrightarrow{x})\right] \right\},\nonumber
\end{eqnarray}
 where $v$ is an interaction potential, and for calculating 
thermodinamically relevant quantities one must consider
$\overrightarrow{x}_1=\overrightarrow{x}_2$ [c.f., Eq.~\eqref{eq:Toll}].

There is no doubt that this formulation of Statistical Mechanics holds.
 However, it turns out to be impossible to understand beyond formal level. 
Two issues arise. First, the Kelvin temperature makes a poor coordinate 
because of the existence of absolute zero which gives a preferred origin. 
Second, the particular way the above path integral depends on $\beta$ makes 
impossible to express a microcanonical ensemble for Statistical Mechanics in 
terms of path integrals, as we expect. These are the main reasons why a new 
formulation of Statistical Mechanics in terms of 
path integrals may be more appropriate. In the remaining of this section, we 
develop our new theory in the {\it nonrelativistic} limit 
(i.e., $|d\overrightarrow{x} /dx^5|\ll 1|$) looking for a Schr\"odinger-like 
equation that will facilitate physical interpretation.\\

Consider $x^5=cu$; so we build a $u$ coordinate with dimension of time (with 
physical meaning of mechanical action in Quantum Mechanics). We define a new 
mass $M$, for which $\Lambda$ plays role of Compton wavelength
\begin{eqnarray}
\label{eq:Lambda}
\Lambda=\frac{\hbar}{Mc}.
\end{eqnarray}
We use the same 5D metric from Sec.~\ref{sec:CanQM} where we made the 
connection
with traditional Quantum Mechanics. It corresponds to the situation of weak
gravitational field (i.e., $\tilde{g}_{ij}=\delta_{ij}$, $\tilde{g}_{0i}=0$,
and $\tilde{g}_{00}=-1-2V/(mc^2)$, with $|2V/(mc^2)|\ll 1$). In addition, we
assume that the electromagnetic field $(A_0, A_i)$ is also weak
(i.e., $|qA_0/c^2|\ll 1$, and  $|qA_i/c^2|\ll 1$), and $\Phi^2\approx 1$. 
Identifying the two alternative parametrizations of the 5D metric given by 
Eqs.~\eqref{eq:fol1} and \eqref{eq:fol2}, straightforward calculations in the 
first order in $2V/(mc^2)$,  $qA_0/c^2$, and $qA_i/c^2$ yield
\begin{eqnarray}
\label{eq:nnms}
\begin{array}{lll}
\phi^2\approx 1+2V/(mc^2),\quad & a_5\approx -qA_0/Q,\quad & a_i\approx 0,\\
G_{ij}\approx g_{ij},\quad & G_{5j}\approx qA_j/c^2,\quad & G_{55}\approx 1.
\end{array}
\end{eqnarray}
 With Eqs.~\eqref{eq:nnms}, the first order of the {\it nonrelativistic} 
expansion (i.e., $|d\overrightarrow{x} /du| \ll c$) of $d_+$ becomes
\begin{gather}
d_{+}\approx c(u_2-u_1)+
c\int^2_1 du
\left[ \frac{1}{2c^2}\left|\frac
{d\overrightarrow{x}}{du} \right|^2 +\frac{q}{c^3}
 \frac{d\overrightarrow{x}}{du}\overrightarrow{A}+\frac{q}{c^2}A_0 
-\frac{V}{mc^2}\right].
\end{gather}

We need an interpretation of the nonrelativistic $u$ in terms of the 
relativistic frame just to clarify what we mean by breaking the 4D covariance 
in the case of Statistical Mechanics. In the case of Quantum Mechanics, the 
nonrelativistic time comes from the relativistic coordinate time. The 
coordinate time takes the place of proper time under the assumption that, in 
the nonrelativistic limit, it makes a good parameter of the time-like path we 
consider. The nonrelativistic approximation $|d\overrightarrow{x} /dt| \ll c$ 
may hold everywhere along a quantum event, with the exception of the special 
point where the path in the future light cone is continued by path in the past
 light cone. A similar interpretation of the nonrelativistic $u$ for 
Statistical Mechanics is not appropriate because the nonrelativistic 
approximation would not hold everywhere along a statistical event (there are 
no causal cones in this case). This calls for a different interpretation of 
the nonrelativistic $u$. The nonrelativistic $u$ comes from the relativistic 
proper time of the statistical loops. We consider the proper time $u'$ of a 
statistical loop to vary along every loop from $u'_1$ to $u'_2$ ($u'_1<u'_2$). 
By breaking 4D covariance in the Riemannian manifold we understand changing 
coordinates from $(\overrightarrow{x},cu)$ to $(\overrightarrow{x},cu')$. This
 transformation opens 4D loops into 3D loops and makes it possible for the 
nonrelativistic approximation $|d\overrightarrow{x} /du'| \ll c$ to hold 
everywhere for some statistical loops. We give the nonrelativistic $u$ the 
physical meaning of the proper time $u' $ which for our version of Statistical
 Mechanics has interpretation of physical time.

The nonrelativistic approximation of the canonical sum over paths
$k_{+}$ is
\begin{eqnarray}
\label{eq:smcansum}
k_{+}(\Lambda^{-1},2,1)=e^{-\Lambda^{-1} c(u_2-u_1)}
\int [d^3x]\exp \left\{ -\frac{1}{\hbar}\int^2_1 du \;
l_{\rm nr}\left( \frac{d\overrightarrow{x}}{du},\overrightarrow{x},u \right)
\right\},\nonumber
\end{eqnarray}
where
\begin{eqnarray}
l_{\rm nr}\left( \frac{d\overrightarrow{x}}{du},\overrightarrow{x},u \right)=
\frac{M}{2}\left|\frac{d\overrightarrow{x}}{du} \right|^2+\frac{Mq}{c}
\overrightarrow{A}\frac{d\overrightarrow{x}}{du}+
(Mq)A_0-\frac{M}{m}V,\nonumber
\end{eqnarray}
and $k_{+}$ is the propagator of a Schr\"{o}dinger-like equation
\begin{eqnarray}
\label{eq:FP}
-\hbar\frac{\partial k_+}{\partial u}=\frac{-1}{2M}
\left[-\hbar\nabla+\frac{Mq}{c}\overrightarrow{A}\right]^2k_+
+(Mq)A_0k_+-
\frac{M}{m}Vk_++Mc^2k_+.
\end{eqnarray}
In order to make the physical interpretation of $M$ and $\Lambda$ transparent, 
consider \eqref{eq:FP} in the case where $\overrightarrow{A}$, $A_0$ and $V$ 
vanish
\begin{eqnarray}
\label{eq:FP2}
\frac{\partial k_+}{\partial u}=\frac{c\Lambda}{2} \nabla^2 k_+-
\frac{c}{\Lambda}k_+.
\end{eqnarray}
We interpret \eqref{eq:FP2} as a Fokker-Planck equation for Brownian motion 
which is the microscopical substrate of Thermodynamics. Therefore, 
$\Lambda=2D/c$, where $D$ is the diffusion constant of the Brownian motion 
\cite{FND}. The theory of Brownian motion predicts $D=1/(\beta\zeta)$, where 
$\zeta$ is a drag coefficient (i.e., from a friction force 
$F_f=-\zeta\, dx/du$ in the Langevin equation; see for example \cite{Coff96}).
 The drag coefficient $\zeta$ can also be writen as $\zeta=m\gamma$, where 
$\gamma$ is called friction constant, and $\gamma^{-1}$ is called 
{\it friction time}. Combining $D=1/(\beta\zeta)$ and $\Lambda=2D/c$ yields
\begin{eqnarray}
\label{eq:lambda1}
\Lambda^{-1}=\beta\, \zeta c/2.
\end{eqnarray}
We know from the traditional Statistical Mechanics that the concept of
temperature (i.e., $\beta$), and not the one of diffusion (i.e., $\beta\zeta$)
characterizes thermal equilibrium. Therefore, for a specific canonical 
ensemble, 
$\zeta$ must be  a constant \cite{FN5}. With this in mind, $\Lambda^{-1}$ does 
characterize thermal equilibrium, as anticipated in the introduction of 
this section.

\begin{figure}\begin{center}
\mbox{\epsfig{file=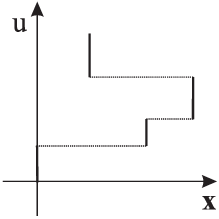,width=60mm}}
\caption{Space-{\it time} picture of Brownian motion. The vertical solid lines 
represent the propagation of the particle, and the horizontal dotted lines 
represent the perturbations of the measurements. Since space-time diagrams do 
not belong to Quantum Mechanics but rather to its classical limit (where the 
fluctuations go to zero), this picture should be thought in the limit of small 
temperatures.}
\label{fig:fig2.5}
\end{center}\end{figure}

From \eqref{eq:Lambda} and \eqref{eq:lambda1} we get an explicit expression
for the mass $M$ of a {\it quantum} particle described by Eq.~\eqref{eq:FP}
\begin{eqnarray}
\label{eq:M}
 M=\frac{\hbar\gamma}{2k_BT}\,m.
\end{eqnarray}
We consider \eqref{eq:FP} not only for the canonical path integral but also
for {\it wavefunctions}. For particles described by such wavefunctions, the 
initial and final positions of propagation always coincide [c.f., 
Eq.~\eqref{eq:Toll}]. A measurement of position yields one of the eigenvalues 
of the position operator. The probability of measuring such an eigenvalue is 
given by
the coefficient of the corresponding eigenvector in the expansion of the 
wavefunction in the position eigenbasis. (Only wavefunctions with all such 
coefficients being positive are given physical meaning as corresponding to 
real particles.) The wavefunction is reduced in the process of measurement 
such that, after the measurement, it equals the eigenvector corresponding to 
the measured eigenvalue. Brownian motion can be thought as a set of such 
measurements in which the particle's wavefunction is reduced in every 
measurement process \cite{Brown}. 
A space-{\it time} diagram of Brownian motion is depicted in 
Fig.~\ref{fig:fig2.5}.
Since space-time diagrams do not belong to quantum physics, but rather to its 
classical limit (where quantum fluctuations tend to zero), this picture 
corresponds to the limit of small temperatures. The universe line has vertical 
segments (solid lines) corresponding to the particle propagation interrupted 
by horizontal segments (dotted lines) corresponding to the perturbations of 
the measurements. Alternatively, one can consider the space and time 
picture of the Newtonian Mechanics provided by the Langevin equation, which 
delivers the same physics in a different interpretation.  
The temperature-dependent stochastic force of the Langevin picture of the
Brownian motion corresponds to the fact that a 4D observer permanently 
measures the statistical particle.\\

We now consider Eq.~\eqref{eq:FP} in the case where all operators in the RHS 
are $u$ independent. The propagator of the differential equation between
$1^4=(\overrightarrow{x},0)$ and $2^4=(\overrightarrow{x},cu)$ can be written as
\begin{eqnarray}
\label{eq:mySM}
k_{+}(\Lambda^{-1},\overrightarrow{x},u)=\sum_a \phi_a^*(\overrightarrow{x})
\phi_a(\overrightarrow{x})e^{-e_a u/\hbar},
\end{eqnarray}
where
\begin{eqnarray}
e_a\phi_a=\frac{-1}{2M}
\left[-\hbar\nabla+\frac{Mq}{c}\overrightarrow{A}\right]^2\phi_a
+(Mq)A_0\phi_a-\frac{M}{m}V\phi_a+Mc^2\phi_a\nonumber,
\end{eqnarray}
resembling the setup of the traditional Statistical Mechanics. One essential 
question remains. We have to find a physical interpretation of $u$ in terms of
4D experimentally accessible quantities. None of the results of the 
traditional Statistical Mechanics contains a mysterious $u$ parameter. The 
fact that the operators in the RHS of \eqref{eq:FP} are now $u$-independent 
(i.e., $x_5$-independent) implies that the statistical particle has a 
well-defined passive mechanical mass. How do we measure $u$ for such a 
physical system under the conditions that we cannot perceive the fifth 
coordinate? Our answer is that 
$u$ has to be proportional with the momentum along the fifth coordinate
$u\propto m$ \cite{FN6}. We can get $m$ by measuring the mass of the whole
thermodynamical ensemble as if it were a {\it single} mechanical particle and 
then divide by the number of particles in the ensemble. This is justified by 
the fact that an ensemble of $N$ noninteracting particles propagating a 
distance $u$ along the
fifth coordinate is equivalent, in terms of suitably defined observables, to
a single particle propagating a distance $Nu$ (Ergodic Principle). Since $u$ 
has meaning of physical time, playing role of action in Statistical Mechanics,
we request $u$ to be {\it quantized} for a particle with definite mass 
(mechanical action is thought to be quantized 
in Quantum Mechanics). We postulate that for the nonrelativistic limit of our
Statistical Mechanics,  a {\it quanta of physical time} for a particle with 
mass $m$ is double its friction time \cite{FNq}
\begin{eqnarray}
\label{eq:u}
u=2m/\zeta=2/\gamma\nonumber.
\end{eqnarray}
Consider now the case of a particle propagating a single quanta of physical time. 
(The case of many quanta propagation can be reduced to this using the Ergodic 
Principle.) Assuming that $V$, $\overrightarrow{A}$ and $A_0$ all 
vanish, 
$e_a$ and $E_a$ scale with $Mc^2$ and $mc^2$, respectively, and we obtain 
\begin{eqnarray}
\label{eq:lambda2}
e_au/\hbar=E_a\beta\nonumber,
\end{eqnarray}
which shows equivalence between our new formulation of Statistical Mechanics 
and Feynman's, for an ensemble of free particles; compare Eqs.~\eqref{eq:Toll}
and \eqref{eq:mySM}. If $V$, $\overrightarrow{A}$ and $A_0$ depend only on the 
elementary charge $e$, Planck's constant $\hbar$, and the speed of light $c$, 
and no other energy scale than $Mc^2$ is introduced by additional dimensional 
constants, we always have that
\begin{eqnarray}
\label{eq:lambda2_}
e_au/\hbar\propto \beta\nonumber.
\end{eqnarray}\\

The most important contribution to the canonical sum over paths in Statistical 
Mechanics is given by paths with local minimum length. These trajectories can 
be interpreted in the context of a new 4D mechanics which we call 
{\it Zero Kelvin Mechanics}. This formalism would eventually prove to be a 
useful 
approximation in the limit $\Lambda\rightarrow 0$. The fundamental equation of 
this mechanics is
\begin{eqnarray}
\label{eq:sst}
\delta d_{+}=0,
\end{eqnarray}
where $d_{+}$ is given by Eq.~\eqref{eq:sd}, and has meaning of physical time.
The extremum condition of the propagation time between two points in space is 
known to physics as the Fermat Principle, which is the founding principle of 
the Geometrical Optics. We get the geometrical physics of light rays from the 
generalized Fermat Principle \eqref{eq:sst} if we remember that 4D photons 
satisfy 
$dx^5=0$, and also demand $Q=0$ [$qA_0=0$ and $a_5(x^\mu)\neq 0$ imply $Q=0$].

Generally, the Lagrangian corresponding to the $d_+$ functional of paths is
 \begin{eqnarray}
l_+(\dot{x}^\mu,x^\mu)=
\sqrt{\tilde{G}_{\mu\nu}\dot{x}^\mu\dot{x}^\nu}-
\frac{Q}{c^2}a_\rho\dot{x}^\rho,~\mbox{where }
\dot{x}^\rho=\frac{dx^\rho}{d\tau}.\nonumber
\end{eqnarray}
The Euler-Lagrange equation of the Lagrangian $l_+$ with the choice of 
parametrization $d\tau=\sqrt{\tilde G_{\mu\nu}dx^\mu dx^\nu}$ is
\begin{eqnarray}
\label{eq:el2}
\dot{x}^\mu\tilde{\triangledown}_\mu \dot{x}^\tau+
\frac{Q}{c^2}\tilde{G}^{\tau\sigma}f_{\sigma\rho}\dot{x}^\rho=0,
\end{eqnarray}
where $\tilde\triangledown_\mu$ is the covariant derivative of the metric
$\tilde{G}^{\mu\nu}$, and $f_{\sigma\rho}=\partial_\sigma a_\rho-
\partial_\rho a_\sigma$. Writing the 5D null path element as
\begin{eqnarray}
\label{eq:last}
\left(\dot{x}^0+\frac{Q}{c^2}a_\mu\dot{x}^\mu \right)^2=1,
\end{eqnarray}
we have Eqs. \eqref{eq:el2} and \eqref{eq:last} representing the 5D null 
geodesic 
equations in a parametrization that breaks 5D covariance. From the fact that
the 5D manifold has a Killing field $\Xi^A=(1,0,0,0,0)$ due to the invariance
 of the metric to translations along $x^0$, there is a conserved quantity
 along the 5D null geodesic
\begin{eqnarray}
\bar M\equiv\frac{h_{AB}}{\phi^2}\,\xi^A\frac{dx^B}{d\Sigma}
=\frac{dx_0}{d\Sigma},\nonumber
\end{eqnarray}
with $\Sigma$ being the affine parameter of the 5D null geodesic of 
$h_{AB}/\phi^2$. The existence of $\bar M$ as a constant of motion 
corresponds to the fact that $M$ (i.e., the Kelvin temperature) is a 
{\it constant of motion} in 
the limit that statistical fluctuations go to zero in the canonical ensemble
of Statistical Mechanics.

In contrast to traditional mechanics, this new theory does not have causal 
cones. Besides the nonrelativistic approximation 
(i.e., $|d\overrightarrow{x}/du|\ll c$),
corresponding to particles close to being at rest, we may also consider a 
{\it superrelativistic} approximation (i.e., $|d\overrightarrow{x}/du|\gg c$)
for particles in the relativistic domain of the traditional mechanics 
\cite{FNsr}. 

\section{The Klein-Gordon Equation}
\label{sec:KG}

In this section, we investigate the problem of counting 5D null paths for the
metric $h_{AB}=diag(-1,1,1,1,1)$. Firstly, we generalize the sum over 5D null 
paths to a sum over 5D paths with a certain length ${\cal D}$
\begin{eqnarray}
{\cal R}({\cal D},2,1)=\int_{{\cal D}=const.}[d^5x].\nonumber
\end{eqnarray}
Secondly, we perform a Wick rotation of the time coordinate 
$x^0\rightarrow ix^0$.
This eliminates the causal structure of the 5D Lorentzian manifold and
transforms it into a 5D Rienmannian manifold. Thirdly, we perform a Laplace
transform of the path integral ${\cal R}({\cal D},2,1)$ with respect to
${\cal D}$, which we denote ${\cal K}({\cal L}^{-1},2,1)$
\begin{eqnarray}
{\cal K}({\cal L}^{-1},2,1)=\int [d^5x]\exp\left(-{\cal L}^{-1}\int_1^2
\sqrt{\delta_{AB}dx^Adx^B}\right).\nonumber
\end{eqnarray}
${\cal K}$ satisfies the following selfconsistency relation \cite{fn3}
\begin{eqnarray}
\label{eq:Ks}
{\cal K}({\cal L}^{-1},2,1)=\int d^5x^{(3)}{\cal K}({\cal L}^{-1},2,3)
{\cal K}({\cal L}^{-1},3,1).
\end{eqnarray}
We can derive a differential equation for the canonical propagator 
${\cal K}({\cal L}^{-1},2,1)$ using \eqref{eq:Ks} for two points 2 and 3 
close to each other (i.e., $2\equiv x^A$ and $3\equiv x^A-\eta^A$, with 
$\delta_{AB}\eta^A\eta^B$ small)
 \begin{eqnarray}
{\cal K}({\cal L}^{-1},x^A,1)=\int d^5\!\eta\;
{\cal K}({\cal L}^{-1},x^A-\eta^A,1)
\;{\cal A} \exp\left(-{\cal L}^{-1}\sqrt{\delta_{AB}\eta^A\eta^B}\right).
\nonumber
\end{eqnarray}
We expand ${\cal K}({\cal L}^{-1},x^A-\eta^A,1)$ in series with respect to 
$\eta^A$ up to the second order, and remark that the linear term in $\eta^A$ 
vanishes by integration
\begin{eqnarray}
{\cal K}({\cal L}^{-1},x^A,1)={\cal K}({\cal L}^{-1},x^A,1)\left[{\cal A}
\int d^5\eta\exp\left(-{\cal L}^{-1}\sqrt{\delta_{AB}\eta^A\eta^B}\right)
\right]+\nonumber\\ +\delta^{AB}
\partial_A\partial_B {\cal K}({\cal L}^{-1},x^A,1)\left[{\cal A}
\int d^5\!\eta\;\delta_{AB}\,\eta^A\eta^B
\exp\left(-{\cal L}^{-1}\sqrt{\delta_{AB}\eta^A\eta^B}\right)\right].\nonumber
\end{eqnarray}
We choose the normalization constant ${\cal A}^{-1}=\int d^5\eta\exp\left(
-{\cal L}^{-1}\sqrt{\delta_{AB}\eta^A\eta^B}\right)$ so that the zeroth order 
of the expansion in $\eta^A$ in the RHS cancels exactly with the LHS. An 
inverse Wick rotation of the resulting partial differential equation for 
${\cal K}$ yields
\begin{eqnarray}
\eta^{\mu\nu}\partial_\mu\partial_\nu\,{\cal K}
+\partial^5\partial_5\,{\cal K}=0,
\nonumber
\end{eqnarray}
where $\eta^{\mu\nu}=diag(-1,1,1,1)$. Finally, a Fourier transformation with 
respect to $x^5$ yields the Klein-Gordon equation 
\begin{eqnarray}
\eta^{\mu\nu}\partial_\mu\partial_\nu \, {\cal K}-
\left(\frac{mc}{\hbar}\right)^2 \! {\cal K}=0.\nonumber
\end{eqnarray}
The Klein-Gordon equation is obtained in the literature from a Lagrangian 
quadratic in $dx^\mu/d\tau$ \cite{Feyn48,Gros95,Gros96} which does not have a 
geometrical meaning.

\section{Local Structure of the 5D Space-Time. 5D Special Relativity}
\label{sec:5DST}
\begin{figure}\begin{center}
\mbox{\epsfig{file=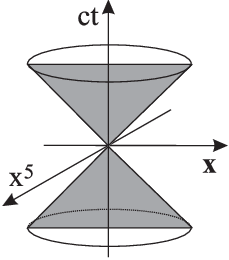,width=60mm}}
\caption{5D causal cone. Projecting its surface in four dimensions yields the 
surface and the interior of the 4D causal cone.} 
\label{fig:fig3}
\end{center}\end{figure}

The 5D space-time has the status of a space-time in General Relativity. In the 
case of Quantum Mechanics, the relation between local 5D causality and local 
4D causality is described by Fig.~\ref{fig:fig3}. Only events on the surface 
of the same 5D causal cone can be connected by a classical trajectory. The 
surface of the local 5D causal cone is isomorphic with a de Sitter space-time 
with zero cosmological constant. Projected in four dimensions, it becomes the 
surface and the interior of the 4D local causal cone. A succession of 4D 
events inside the local 4D cone corresponds to a succession of 5D events on 
the surface of the local
5D cone with $\Delta x^5$ being either positive or negative (i.e., positive or 
negative particle mass). The local geometry of the 5D space-time is 
Minkovski-like and its isometry group is O(4,1). This group is known as the de
 Sitter group or, 
for other purposes, as the quantum number group of the hydrogen atom. The 
irreducible representations of O(4,1) have been completely classified 
\cite{Schw71}. O(4,1) can be described by SO(4,1) and two discrete 
transformations:
time reversal T (i.e., T transforms $x^0\rightarrow -x^0$) and CP, where P is 
parity (i.e., P transforms $\overrightarrow{x}\rightarrow-\overrightarrow{x}$),
and C transforms $x^5\rightarrow -x^5$. Applying C in the context of Quantum
Mechanics corresponds to changing the direction of foliation from $x^5$ to 
$-x^5$, which is equivalent to changing the sign of the shift of the foliation 
$N_\mu=-qA_\mu/c^2$. Therefore, C is equivalent to a specific charge 
conjugation 
$q\rightarrow -q$. C can also be thought as leaving $q$ unchanged, instead 
transforming the mass $m$ of the quantum particle to $-m$.

The foliated physical interpretation breaks the O(4,1) symmetry into 
O(3,1)$\bigotimes$U(1) for Quantum Mechanics, and into O(4)$\bigotimes$D(1) for 
Statistical Mechanics, where D(1) denotes the dilatation group in one 
dimension. In contrast with the symmetry breaking in the Quantum Field Theory,
these are not proper physical phenomea, instead they are due to how the 4D observer
foliates the 5D manifold to four dimensions.\\

\begin{figure}\begin{center}
\mbox{\epsfig{file=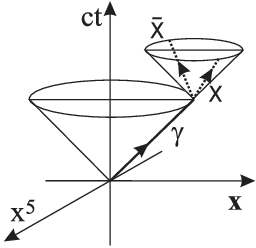,width=60mm}}
\caption{Creation of particle-antiparticle pair in 5D space-time.}
\label{fig:fig4}
\end{center}\end{figure}

It is both natural and useful to introduce a Special Relativity for the 5D 
Optics.
Since this requests a 5D Minkovski-like metric, there is no gravitational or 
electromagnetic interaction in this theory. The only type of interaction is 
collision. There are two types of particles in 4D Special Relativity: photons 
with
null 4-momentum, and massive particles with 4-momentum of magnitude $mc$.
In this theory, we only have 5D photons (i.e., on-shell particles) with null 
5-momentum
\begin{eqnarray}
p^A=(p^\mu,mc).\nonumber
\end{eqnarray}
We postulate conservation of 5-momentum in a collision process. An example of 
a phenomenon that gains understanding from these considerations is 
particle-antiparticle pair creation $\gamma\rightarrow X +\bar X$. (We ignore 
effects related to the charges and the spins of the particles.) A 4D photon 
$\gamma$ with wavevector $\overrightarrow{k}$ and energy $\hbar \omega$ 
collides with another 4D photon with arbitrary small energy. Conservation of 
5-momentum writes
\begin{eqnarray}
(\hbar \omega/c, \hbar \overrightarrow{k},0)=
(E_X/c,\overrightarrow{p}_X,m_Xc)+(E_{\bar X}/c,\overrightarrow{p}_{\bar X},
m_{\bar X}c).\nonumber
\end{eqnarray}
Therefore, we get $m_{\bar X}=-m_{X}$, and the conservation of 4-momentum. 
After collision, the 4D photons are perceived in four dimensions as a 
particle-antiparticle pair (see Fig.~\ref{fig:fig4}). This {\it rotation} of a
4D photon pair into a particle-antiparticle pair cannot be described in the 
context 
of Quantum Mechanics because it implies a noncylindrical transformation of 
coordinates. In 5D Special Relativity, these transformations contain either a 
boost along $x^5$ or a rotation of $x^5$.\\

For the 5D flat metric, both the quantum and the statistical interpretations 
are possible. For a particle with well defined mass, at thermodynamical 
equilibrium (i.e., macroscopically at rest), we can write the null 5-momentum 
in two equivalent forms (up to a constant factor)
\begin{eqnarray}
\label{eq:pqm}
p^A&\propto &(E/c,0,mc),\\
\label{eq:psm}
p^A&\propto &(Mc,0,e/c).
\end{eqnarray}
If we believe that Eqs.~\eqref{eq:pqm} and \eqref{eq:psm} represent different 
pictures of the same 5D reality, we have
\begin{eqnarray}
 e&\propto &E,\nonumber\\
\label{eq:M&m}
 M&\propto &m.
\end{eqnarray}
Remark that the local transformation group that leaves this interpretation intact is 
the group of spatial rotations $O(3)$, because of the covariance breaking in 
Quantum and Statistical Mechanics. Equation \eqref{eq:M&m} can be interpreted 
in the following way. Assume that we have a particle macroscopically at rest, 
subject to 
periodic measurements. It is ambiguous whether what we measure are the quantum 
fluctuations of a particle with passive mass $m$, or the Brownian motion of a 
particle at temperature $T$. If many particles of mass $m$ formed a canonical 
statistical ensemble, it would have temperature $T\propto \hbar\zeta/(2mk_B)$ 
[c.f., Eqs.~\eqref{eq:M} and \eqref{eq:M&m}]. However, 5D Special Relativity 
being a classical limit (in the sense that fluctuations go to zero), this 
relation holds only as $m\rightarrow\infty$ and $T\rightarrow 0$.

\section{Further Discussions and Conclusions}
\label{sec:x3}
Based on the results of Secs.~\ref{sec:QM} and \ref{sec:5DST}, we are now in 
the position to discuss the foliation of $h^{AB}$ along a proper space 
dimension, $x^3$. We assume that $h^{AB}$ is independent of $x^3$. The 
mathematical structure
of the formalism is similar to that of Sec.~\ref{sec:QM}. For a physical 
interpretation, consider the situation of weak gravitational field (i.e., 
$\tilde{g}_{ij}=\delta_{ij}$, $\tilde{g}_{0i}=0$, and 
$\tilde{g}_{00}=-1-2V/(mc^2)$, with $|2V/(mc^2)|\ll 1$), $A_i=0$, 
$|qA_0/c^2|\ll 1$, and $\Phi^2\approx 1$. We restrict to first order 
calculations in $2V/(mc^2)$ and $qA_0/c^2$. Solving the 5D null path element 
in $dx^3$ and then integrating yields
\begin{eqnarray}
\Delta x^3\equiv \Delta z\approx\pm\int_1^2\left[\sqrt{
c^2dt^2-dx^2-dy^2-(dx^5)^2}-\frac{q}{c^2}A_0dx^5+\frac{V}{mc^2}cdt\right]
\nonumber,
\end{eqnarray}
where $c\,t\equiv x^0$, $x\equiv x^1$, $y\equiv x^2$, and $z\equiv x^3$. We 
assume that $(dx^{1,2,5}/dt)^2\ll c^2$, and we expand $\Delta x^3$ in this
{\it nonrelativistic} limit. This corresponds to the ultrarelativistic limit 
of the traditional 4D mechanics [i.e., $p_{x,y}^2\ll (E/c)^2$, and 
$m^2c^2\ll (E/c)^2$]. We denote the canonical path integrals of the foliation 
along $x^3$ by $K_{z\pm}$. They are propagators of Schr\"odinger-like 
equations 
\begin{eqnarray}
\label{eq:l_Schr}
\pm\frac{\hbar}{i}\frac{\partial K_{z\pm}}{\partial t}=\frac{c}{2p_z}
\left[ \left( \frac{\hbar}{i}\frac{\partial}{\partial x}\right)^2+
\left(\frac{\hbar}{i}\frac{\partial}{\partial y}\right)^2+
\left(\frac{\hbar}{i}\frac{\partial}{\partial x^5}+\frac{p_z}{c^2}qA_0
\right)^2\right]K_{z\pm}\nonumber\\+\frac{p_z}{mc}VK_{z\pm}+p_zcK_{z\pm},
\end{eqnarray}
where $p_z$ is the momentum along the z-direction and constant of motion in 
this 
ultrarelativistic approximation.  Under the assumption that $V$ and $A_0$ are 
independent of $x^5$, a Fourier transform of \eqref{eq:l_Schr} with respect to 
$x^5$ yields
\begin{eqnarray}
\label{eq:L_Schr}
\pm\frac{\hbar}{i}\frac{\partial K_{z\pm}}{\partial t}=\frac{c}{2p_z}
\left[\left(\frac{\hbar}{i}\frac{\partial}{\partial x}\right)^2+
\left(\frac{\hbar}{i}\frac{\partial}{\partial y}\right)^2\right]K_{z\pm}+
\nonumber\\\left[ \frac{c}{2p_z}
\left(mc+\frac{p_z}{c^2}qA_0
\right)^2+\frac{p_z}{mc}V+p_zc\right]K_{z\pm}.
\end{eqnarray}
We interpret \eqref{eq:L_Schr} as a (2+1)D Schr\"odinger equation for the motion
of an ultrarelativistic quantum particle with $p_z^2\gg m^2c^2$ transverse to 
the z-direction. The fields in the RHS of \eqref{eq:L_Schr} must allow $p_z$ 
to be an approximate constant of motion, $p_z^2\gg p_{x,y}^2$, and 
$p_z^2\gg m^2c^2$, for 
this approximation of the 5D Quantum Optics formalism to be valid. We believe 
that Eq.~\eqref{eq:l_Schr} could easily be tested with the current accelerator 
technology \cite{fn_l}. 
If $V$ and $A_0$ in Eq.~\eqref{eq:l_Schr} depend on $x^5$, then the 
passive mass of the quantum particle varies with time. Thus, according to this
 theory, the mass variation of ultrarelativistic particles (e.g., neutrinos) 
probes the $x^5$ dependence of the 5D metric.\\

In this paper we have investigated a 5D space-time geometry equipped with a 
quantum principle for 5D spinless and masless particle propagation. We have 
given physical interpretation to 4D foliations with corresponding symmetries. 
We have shown that we can regard the foliation of the fifth dimension (which 
escapes direct perception) as a formulation of Quantum Mechanics, while the 
foliation of the time dimension can be interpreted in the terms of Statistical
 Mechanics. The foliation along a proper space-like dimension yields a 
description of a spinless ultrarelativistic particle. We have derived 
Schr\"odinger-like equations for the nonrelativistic approximations of the 
canonical path integrals corresponding to the different foliations. We have 
been particularly interested in the case of weak gravitational and 
electromagnetic fields, but the mathematical theory of path 
integrals for quadratic lagrangeians in curved manifolds (see for example 
\cite{Khan,Gros96,Langouche,Chaichian} and references therein) may allow for 
deriving Schr\"odinger-like equations in more general cases (exact solutions 
of the field 
equations). We have also introduced a 5D Special Relativity completing our 
interpretation of the 5D geometry. In conclusion, the interpretation of the 5D 
geometry with respect to 4D physical concepts depends on the particular 4D 
foliation of the 5D manifold. We emphasize that, in general, these 
interpretations are mutually exclusive. For a 4D interpretation to hold, one 
must restrict to cylindrical transformations of coordinates (in accord to the
coordinate independence).

However, a general 5D metric is not expected to posses the symmetry that is 
necessary in order to apply our 4D interpretations of the 5D geometry.  
Nevertheless, these interpretations may approximately hold as the requested 
symmetries approximately hold, leading to physical theories with limited 
validity.  In fact, since {\it vacuum} is an abstraction never to be reached 
in practice, Quantum Mechanics cannot claim to be a ultimate theory; nor can 
Statistical Mechanics since {\it thermal equilibrium} is reached in an 
infinite amount of time.

\newpage

\begin{thebibliography}{99}
\bibitem{Wess83} P.S. Wesson, Astron. Astrophys. {\bf 119}, 145 (1983) is
 one of the early publications where Wesson discusses the idea that the fifth
coordinate is proportional to mass.
\bibitem{kkmetric} The 5D Kaluza-Klein metric (see e.g., Ref. 
\cite{Wesson} or Eq. \eqref{eq:fol1} below) is independent on the fifth 
coordinate.
\bibitem{Wessmass} One may argue that it is possible to resolve the 
inconsistency between the interpretation of the fifth dimension and that of the
4D on-shell constraint in Wesson's gravity by using the distinction between
passive and active mass. Indeed, $p^5=mc$ relates to passive mass, while
$x^5=Gm/c^2$ relates to active mass. However, it is then hardly conceivable 
how covariance holds when a coordinate determines source terms in the Einstein
equations!
\bibitem{jacobian} The jacobian $\sqrt{|h|}$ is irrelevant to path counting.
It is introduced just formaly.
\bibitem{cyl}  The class of transformations which preserve 
Eqs.~\eqref{eq:fieldeq2} is more general. Some noncylindrical transformations 
are equivalent to changes of the electromagnatic gauge 
(see Sec.~\ref{sec:CanQM}). Cylindrical transformations, as proposed by Kaluza
 \cite{Kaluza21_}, preserve Eqs.~\eqref{eq:fieldeq2} and the 
electromagnetic gauge.
\bibitem{m=0} In the case where $m=0$ (i.e., $\lambda^{-1}=0$), the canonical
 path integral can be calculated exactly
\begin{eqnarray}
K_{\pm}(\lambda^{-1},2^4,1^4)
=\int [d^4x]\; e^{iD_{\pm}\lambda^{-1}}=\int [d^4x]=(const.)\,.\nonumber
\end{eqnarray}
We obtain the microcanonical path integral by inverse Fourier transform
\begin{eqnarray}
R_{\pm}(D,2^4,1^4)=(const.)\; \delta(D),\nonumber
\end{eqnarray}
confirming that 4D particles with zero rest mass (i.e., 4D photons) propagate
exclusively on 4D null paths.
\bibitem{FN2} This is the original formulation of the Feynman path integral 
for Statistical Mechanics \cite{Feyn65,Feyn72}. Whether $u_1$ is zero or not 
is irrelevant for the predictions of this theory which depend only on the 
difference $u_2-u_1$.
\bibitem{FN6} This is consistent for 4D photons which have 
$u=\Delta x^5/c=0$, and also $m=0$.
\bibitem{FNsr} In the simplified case where we drop the electromagnetic field 
and consider a flat 4D Rienmannian metric, we have
\begin{eqnarray}
\label{eq:smsuprel}
l(\dot{\overrightarrow{x}},\overrightarrow{x},u)=Mc\sqrt{\dot
{\overrightarrow{x}}^2+c^2}=
Mc|\dot{\overrightarrow{x}}|+\frac{Mc^3}{2|\dot
{\overrightarrow{x}}|}+...\,.\nonumber
\end{eqnarray}
The first term of the expansion is the well-known optical Lagrangian in an 
isotropic medium. Considering the second term, it can be shown 
straightfowardly that this new mechanics obeys the law of inertia. Thus, in 
general, for minimum order approximation of the dynamics we may keep only the 
first order kinetic and potential terms of the Lagrangian and apply the law of
 inertia.
\bibitem{fn3} The integration over the intermediate events 3 is not 
constrained now by causal structure.
\bibitem{fn_l} As an example, consider the case where $V=0$ and $A_0$
describes a perturbing electric field perpendicular to the z-direction, and 
periodic in time. Using the time-dependent perturbation theory of the 
Schr\"odinger equation, one can show that the kinetic energy of the 
ultrarelativitic particle along the direction of the electric field has a 
diffusive spread proportional with the square amplitude of the electric field.
\end{thebibliography}
\end{document}